\documentclass[twocolumn,english,aps,prl,showpacs]{revtex4}
\usepackage[T1]{fontenc}
\usepackage[latin1]{inputenc}
\usepackage{float}
\usepackage{amsmath}
\usepackage{graphicx}
\usepackage{amssymb}

\makeatletter

\usepackage{babel}
\makeatother
\begin{document}

\preprint{preprint}

\title{Phase diagram  for the Grover algorithm with static imperfections}

\author{A.A.~Pomeransky$^{(a)}$, O.V. Zhirov$^{(b)}$ and
D.L.~Shepelyansky$^{(a)}$}
\affiliation{$^{(a)}$Laboratoire de Physique Th\'eorique, 
UMR 5152 du CNRS, Universit\'e P. Sabatier, 31062 Toulouse Cedex 4, France\\
$^{(b)}$Budker Institute of Nuclear Physics, 630090 Novosibirsk, Russia 
}

\date{March 19, 2004}

\begin{abstract}
We study effects of static inter-qubit interactions on the stability of 
the Grover quantum search algorithm.
Our numerical and analytical results show existence of regular and 
chaotic phases depending on the imperfection
strength $\varepsilon$. 
The critical border $\varepsilon_c$ between two phases drops polynomially 
with the number of qubits $n_q$ as $\varepsilon_c \sim n_q^{-3/2}$. 
In the regular phase $(\varepsilon < \varepsilon_c)$ the algorithm remains 
robust against imperfections showing the efficiency gain $\varepsilon_c / \varepsilon$ 
for $\varepsilon \gtrsim 2^{-n_q/2}$.
In the chaotic phase $(\varepsilon > \varepsilon_c)$ the algorithm
is completely destroyed. 
\end{abstract}
\maketitle

Quantum computations open new perspectives and possibilities for treatment of complex computational
problems in a more efficient way with respect to algorithms based on the classical logic
\cite{Nielsen}. In the quantum computers classical bits are replaced by two-level quantum systems
(qubits) and classical operations with bits are substituted by elementary unitary transformations
(quantum gates). The elementary gates can be reduced to single qubit rotations and controlled two-qubit
operations, e.g. {\it control-NOT} gate \cite{Nielsen}. Combinations of elementary gates allow to 
implement any unitary operation on a quantum register, which for $n_q$ qubits contains exponentially 
many states $N=2^{n_q}$. The two most famous quantum algorithms are the Shor algorithm for integer 
number factorization \cite{Shor94} and the Grover quantum search algorithm \cite{Grov97}. 
The Shor algorithm is exponentially faster than any known classical algorithm, while the Grover
algorithm gives a quadratic speedup. 

In realistic quantum computations the elementary gates are never perfect and
therefore it is very important to analyze the effects of imperfections and quantum errors on 
the algorithm accuracy. A usual model of quantum errors assumes that angles of unitary
rotations fluctuates randomly in time for any qubit in some small interval $\varepsilon$ near 
the exact angle values determined by the ideal algorithm. In this case a realistic quantum
computation remains close to the ideal one up to a number of performed gates  
$N_g \sim 1/\varepsilon^2$. For example, the fidelity $f$ of computation, defined as
a square of scalar product of quantum wavefunctions of ideal and perturbed algorithms, 
remains close to unity if a number of performed gates is smaller than $N_g$. This result
has been established analytically and numerically in extensive studies of various quantum 
algorithms \cite{Cirac95,Paz,Geor01,Song02,Terra03,Bett03,Frahm03}.

Another source of quantum errors comes from internal imperfections generated by residual static
couplings between qubits and one-qubit energy level shifts which fluctuate from one qubit to 
another but remain static in time. These static imperfections may lead to appearance of many-body
quantum chaos, which modifies strongly the hardware properties of realistic quantum computer
\cite{Geor00,Berman,BeneEPJD}.
The effects of static imperfections on the accuracy of quantum computation have been investigated 
on the examples of quantum algorithms for the models of complex quantum dynamics 
\cite{Bene01,Pome04,Terra03,Frahm03}.
As a result a universal law for fidelity decay induced by static imperfections has been 
established \cite{Frahm03} for quantum algorithms simulating dynamics in the regime of 
quantum chaos. 
At the same time it has been realized that the effects of static imperfections for 
dynamics in an integrable regime are not universal and more complicated. 
Therefore it is important
to investigate the effects of static imperfections on an example of the well known
Grover algorithm. First attempt was done recently in 
\cite{Braun02}, but the global picture of the phenomenon remained unclear. In this paper
we present extensive numerical and analytical studies which establish the global stability
diagram of reliable operability of the Grover algorithm.

Let us first outline the key features of the Grover algorithm \cite{Grov97}. 
An unstructured 
database is presented by $N=2^{n_q}$ states of quantum register with $n_q$ qubits: 
$\lbrace\vert x\rangle \rbrace $, $x=0,\ldots,N-1$.  
The searched state $\vert \tau \rangle $ can be identified by {\it oracle} function $g(x)$, defined as
$g(x)=1 $ if $x=\tau$ and $g(x)=0 $ otherwise. The Grover iteration operator $\hat{G} $ is a product
of two operators: $\hat{G}=\hat{D} \hat{O}$. Here the oracle operator $\hat{O}=(-1)^{g(\hat{x})}$
is specific to the searched state $\vert \tau \rangle $, while the diffusion operator $\hat{D}$ is
independent of  $\vert \tau \rangle $: $D_{i i}=-1+\frac{2}{N}$ and 
$ D_{i j}=\frac{2}{N} \;\; (i\neq j)$.
For the initial state $\vert \psi_0 \rangle=\sum_{x=0}^{N-1}\vert x \rangle/\sqrt{N}$, 
$t$ applications of the Grover operator $\hat{G}$ give \cite{Nielsen}:
\begin{equation}
\label{eq:rotGr}
       \vert \psi(t) \rangle=\hat{G}^t\vert \psi_0 \rangle
	   =\sin{((t+1) \omega_G)}\vert \tau \rangle +\cos{((t+1) \omega_G)}\vert \eta \rangle \; ,
\end{equation}
where the Grover frequency $\omega_G=2\arcsin(\sqrt{1/N})$ and
$\vert \eta \rangle=\sum_{x\neq\tau}^{(0\leq x<N)}\vert x \rangle/\sqrt{N-1}$.
Hence, the ideal algorithm
gives a rotation in the 2D plane $(\vert \tau \rangle,\vert \eta \rangle)$.

The implementation of the operator $D$ through the elementary gates requires an 
ancilla qubit. As a result the Hilbert space becomes a sum of two subspaces
$\lbrace \vert x\rangle \rbrace$ and $\lbrace \vert x+N\rangle \rbrace$, which
differ only by a  value of $(n_q+1)$-th qubit. These subspaces are invariant with respect to operators
$O$ and $D$: $O=1-2\vert \tau\rangle\langle\tau\vert-2\vert \tau+N\rangle\langle\tau+N\vert$,
$D=1-2\vert \psi_0\rangle\langle\psi_0\vert-2\vert \psi_1\rangle\langle\psi_1\vert$, where
$\vert \psi_1 \rangle=\sum_{x=0}^{N-1}\vert x+N \rangle/\sqrt{N}$ and $\vert \psi_{0,1}\rangle$
correspond to up/down ancilla states. 
Then $D$ can be represented  as $D=W R W$ \cite{Grov97}, where the transformation 
$W=W_{n_q}\ldots W_k\ldots W_1$ is
composed from $n_q$ one-qubit Hadamard gates $W_k$, and $R$ is
the $n_q$-controlled phase shift defined as 
$R_{i j}=0$ if $i\neq j$, $R_{0 0}=1$ and $R_{i i}=-1$ if $i\neq 0$
 $(i,j=0,\ldots,N-1)$. In turn, this operator can be represented as
$R=W_{n_q}\sigma^x_{n_q-1}\ldots \sigma^x_{1}\wedge_{n_q} \sigma^x_{n_q-1}\ldots \sigma^x_{1}W_{n_q}$,
where $\wedge_{n_q}$ is generalized $n_q$-qubit Toffolli gate, which inverts the $n_q$-th qubit if the first
$n_q-1$ qubits are in the state $ \vert 1\rangle$. The construction of $\wedge_{n_q}$ from $3$-qubit
Toffolli gates with the help of only one auxillary qubit is described in \cite{Bare95}.
As a result the Grover operator G is implemented through $n_g=12n_{tot}-42$ elementary gates including 
one-qubit rotations, control-NOT and Toffolli gates. Here $n_{tot}=n_q+1$ is the total number of qubits.

To study effects of static imperfections on the Grover algorithm we use the model introduced
in \cite{Geor00}.  In this model a quantum computer hardware is described by the Hamiltonian
$H$:
\begin{equation}
    H=\sum_i \frac{\Delta}{2} \sigma_i^z +H_S, \;\;
    H_S=\sum_{i} a_i\sigma_i^z+\sum_{i<j} b_{i j} \sigma^x_i\sigma^x_j .
	\label{eq:Hint}
\end{equation}
Here, $\sigma_i$ are the Pauli matrices for qubits $i$, and $\Delta$ is an average
one-qubit energy spacing. All $n_{tot}$ qubits 
are placed on a rectangular lattice, the second sum in $H_S$ runs over
nearest neighbor qubits with periodic boundary conditions. 
Qubit energy shifts $a_i$ and couplings $b_{ij}$ are randomly and uniformly distributed 
in the intervals $\left[-\alpha,\alpha\right]$ and $\left[-\beta,\beta\right]$, respectively.
Following \cite{Bene01,Terra03,Frahm03,Pome04}
we assume that the average spacing $\Delta$ is compensated by specially applied laser pulses 
so that between subsequent elementary gates the wavefunction evolution is given by the propagator
$U_S=\exp(-i H_S t_g)$. Thus all static errors are expressed via this propagator while the elementary 
gates are taken to be perfect. Appropriate rescaling of parameters $a_i$ and $b_{ij}$
allows to put $t_g=1$ without any loss of generality. We concentrate our studies on the case
 $\alpha=\beta\equiv\varepsilon$ where inter-qubit couplings lead to a developed quantum chaos
\cite{Geor00,Bene01}. 

\begin{figure}[th]
   \centering
   \includegraphics[scale=0.3,angle=90]
   {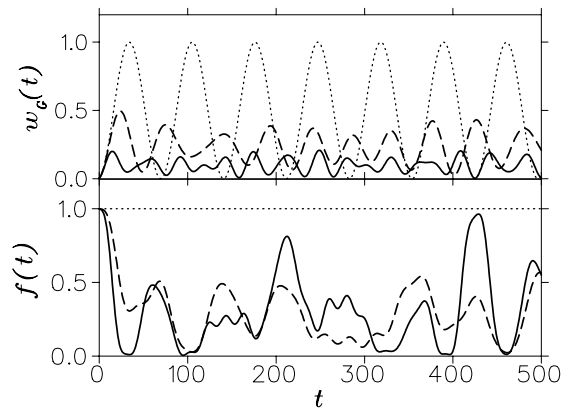}
   \caption{Probability of searched state $w_G(t)$ (top) and fidelity $f(t)$ (bottom)  
   as a function of the iteration step $t$ in the Grover algorithm for $n_{tot}=12$ qubits. 
   Dotted curves show results for the ideal algorithm $(\varepsilon=0)$, dashed and 
   solid curves correspond to imperfection strength $\varepsilon= 4\cdot 10^{-4}$ and 
   $10^{-3}$, respectively.} 
   \label{fig1}
\end{figure}

A typical example of imperfection effects on the accuracy of the Grover algorithm is shown in Fig.\ref{fig1}
for a fixed disorder realization of $H_S$ in (\ref{eq:Hint}) on $3\times 4$ qubit lattice. 
It clearly shows that imperfections suppress the probability 
$w_G$ to find the searched state, where $w_G$ is given by a sum of probabilities of states
$\vert \tau\rangle$ and $\vert \tau+N\rangle$.  
In contrast to the case of time-dependent random quantum errors 
studied in \cite{Song02} in the case of static imperfections the oscillations of probability $w_G$  
do not decrease with time $t$. Another interesting feature is a significant decrease of the 
period of the Grover oscillations compared to the case of ideal algorithm, where 
$T_G=\pi/2\omega_G$. This effect is also absent in the case of random errors.
The fidelity of quantum computation $f(t)$ also shows non-decaying oscillations at large times.
However, in average the maxima of fidelity correspond to minima rather than maxima of probability
$w_G$. Hence, $f(t)$ is not appropriate for tests of the algorithm accuracy.

\begin{figure}[th]
   \centering
   \includegraphics[scale=0.6]{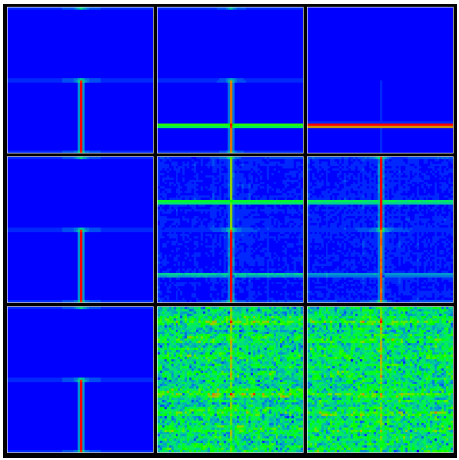}
   \caption{(Color) Evolution of the Husimi function in the Grover algorithm at times $t=0$, 17, and $34$ 
   (from left to right), and for $\varepsilon=0$, 0.001, and 0.008 
   (from top to bottom). The qubit lattice and disorder realization are the same as in Fig.\ref{fig1}.
   The vertical axis shows the computational basis $x=0,\ldots,2N-1$, while the horizontal axis 
   represents the conjugated momentum basis. Density is proportional to color changing from maximum
   (red) to zero (blue).  
   }
   \label{fig2} 
\end{figure}

Following \cite{Hus-Paz} a pictorial presentation of the dynamical evolution in the Grover algorithm can 
be obtained with the help of the Husimi function \cite{Husimi}, which is shown in Fig.\ref{fig2}.
In this presentation the computational basis $x$ can be considered as a coordinate space representation
for the wavefunction $\psi_x(t)$ ($x=0,\ldots,2N-1$), while the conjugated basis obtained by the Fourier
transform corresponds to momentum representation $p$ ($p=-N+1,\ldots,N$). In this way the initial state 
of the Grover algorithm $\vert \psi_0 \rangle$ gives a peaked distribution with $p=0$.
In the ideal algorithm the total probability is distributed between two states  $\vert \tau \rangle$ 
and $\vert \eta \rangle$ (see Eq.(\ref{eq:rotGr})) that gives two orthogonal lines in the phase space
of Husimi function (see Fig.\ref{fig2}, top raw). After the period $T_G\approx 34$ all the probability is 
transferred to the target state $\vert \tau \rangle$ ($w_G\approx 1$). In the presence of
moderate imperfections
the flips of the ancilla qubit become possible that involves into dynamics two additional states. 
As a result 
the probability is mainly distributed over {\it four states} corresponding to four straight lines 
in phase space (Fig.\ref{fig2}, middle raw):
\begin{equation}
  \begin{array}{ll}
      \vert \tau_0\rangle = \vert \tau\rangle  \quad& \vert \tau_1\rangle=\vert \tau+N\rangle\\
      \vert \eta_0\rangle = \vert \eta\rangle  \quad&
      \vert \eta_1 \rangle=\sum_{x\neq\tau}^{(0\leq x<N)}\vert x+N \rangle/\sqrt{N-1}
  \end{array}.
\label{eq:4vect}
\end{equation}
The probability $w_4$ contained in these states is close to unity (in Fig.\ref{fig2} $w_4=0.998$ for 
$\varepsilon=10^{-3}$).
Above certain critical border $\varepsilon_c$ this simple structure is completely washed out
($w_4=6\cdot10^{-4}$), and
the Husimi function shows only random distribution (Fig.\ref{fig2}, bottom raw).

\begin{figure}[th]
   \centering
   \includegraphics[width=80mm,height=70mm]
   {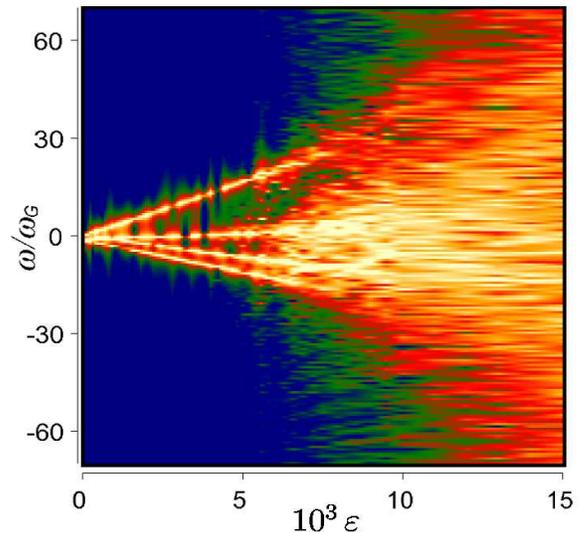}
   \caption{(Color) Phase diagram for the spectral density $S(\omega)$ as a function of imperfection
   strength $\varepsilon$, $n_{tot}=12$, same disorder realization as in Fig.\ref{fig2}.
   Color is proportional to density $S(\omega)$ (yellow for maximum and blue for zero).
   }
    \label{fig3}
\end{figure}

The dominant contribution of these four states can be also seen in spectral density $S(\omega)$ 
of the wavefunction $\psi_x(t)$. This density is defined as: $S(\omega)=\sum_x |a_x(\omega)|^2$, where
$a_x(\omega)=\sum_{t=0}^{T_f} \psi_x(t)\exp(i\omega t)/\sqrt{T_f}$ and $T_f$ is a large time scale
on which the spectrum is determined (we usually used $T_f\approx 5 T_G \gg T_G$). 
The phase diagram of spectral density $S(\omega)$ dependence on the imperfection strength $\varepsilon$
is shown in Fig.\ref{fig3}. Two phases are clearly seen: for $\varepsilon < \varepsilon_c$ the diagram 
contains four lines corresponding to the four states (\ref{eq:4vect}), while for 
$\varepsilon > \varepsilon_c$ these lines are destroyed and the spectrum becomes continuous. These 
phases correspond to the qualitative change of the Husimi distribution shown in Fig.\ref{fig2}. 

\begin{figure}[th]
   \centering
   \includegraphics[width=63mm,height=78mm,angle=90]
   {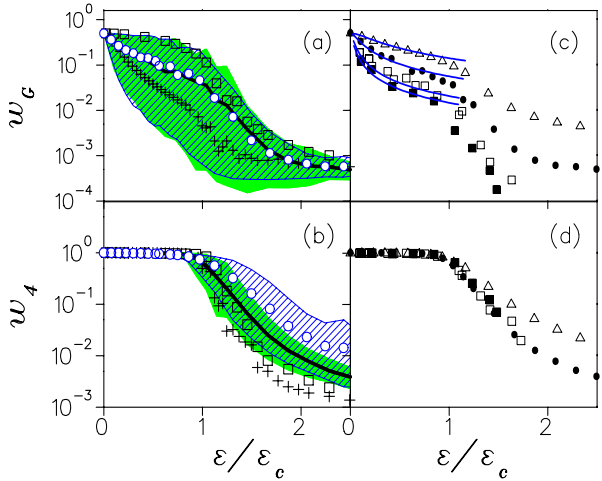}
   \caption{(Color online) Dependence of probabilities $w_G$ (a,c) and $w_4$ (b,d) on rescaled 
   imperfection strength $ \varepsilon/ \varepsilon_c$, with $\varepsilon_c$ from (\ref{eq:eps_c}). 
   For panels (a,b) $n_{tot}=12$, squares and pluses show data for two typical disorder realizations,
   green/grey area shows the region of probability variation for various disorder realizations (see
   text), full thick curves give average dependence $\bar{w}_{G}$, $\bar{w}_{4}$. Dashed area bounded by
   thin curves show the region of probability variation in the single-kick model, open circles give
   the average data in this model with rescaling factor $R=0.56$.        
   Panels (c,d) show $\bar{w}_{G}$, $\bar{w}_{4}$ for $n_{tot}=9$ (triangles), 12 (full circles), 15 
   (open squares) and 16 (full squares). In panel (c) full curves are given by Eq.(\ref{eq:2stMean}) 
   for same  $n_{tot}$ values from top to bottom, $R=0.56$. 
   }
   \label{fig4}
\end{figure}

To study the transition between these phases in a more quantitative way we analyze the dependence
of probabilities $w_G$ and $w_4$ on the imperfection strength $\varepsilon$ for a large number of
disorder realizations in $H_S$ (\ref{eq:Hint}) changing also the number of qubits $n_{tot}$. 
The number of realizations vary from 50 to 1000 depending on $\varepsilon$ and $n_{tot}$.  
Since the frequency of Grover oscillations varies strongly with $\varepsilon$ and disorder we 
average $w_G$ and $w_4$ over a large time interval $T_f$ to suppress fluctuations in time.
The obtained results are summarized in Fig.\ref{fig4}. For a fixed value of $n_{tot}$ the dependence
$w_G(\varepsilon)$ changes strongly from one realization to another (Fig.\ref{fig4}a). In contrast,
the probability $w_4$ remains close to unity being insensitive to variations of disorder up to
$\varepsilon < \varepsilon_c$ (Fig.\ref{fig4}b). Only for $\varepsilon > \varepsilon_c$,    
when $w_4 \ll 1$, it becomes sensitive to disorder. The probabilities averaged over disorder
$\bar{w}_G$ and $\bar{w}_4$ are shown in Fig.\ref{fig4}a,b. They also have a qualitative 
change in behavior near $ \varepsilon_c$, especially $\bar{w}_4$. 
These results confirm the fact that the phase 
transition takes place near some critical $\varepsilon_c$ for an ensemble of disorder realizations.

The value of $\varepsilon_c$ can be obtained from the following estimate. The transition rate induced
by imperfections after one Grover iteration is given by the Fermi golden rule: 
$\Gamma \sim \varepsilon^2 n_g^2 n_{tot}$, where $n_{tot}$ appears due to random contribution of qubit
couplings $\varepsilon$ while $n_g^2$ factor takes into account coherent accumulation of perturbation
on $n_g$ gates used in one iteration (see, e.g. \cite{Frahm03}). In the Grover algorithm the four states
(\ref{eq:4vect}) are separated from all other states by energy gap $\Delta E \sim 1$ 
(it appears due to sign change introduced by operators $O$ and $D$). 
Thus these four states become mixed with all others for 
\begin{equation}
    \varepsilon > \varepsilon_c \approx 1.7 /(n_g\sqrt{n_{tot}})
    \label{eq:eps_c}
\end{equation}  
when  $\Gamma > \Delta E$. Here the numerical factor is obtained from numerical data. The parameter 
dependence is well confirmed by data for $\bar{w}_4$ shown in Fig.\ref{fig4}d.

The variation of averaged Grover probability $\bar{w}_G$ with $\varepsilon$ and $n_{tot}$ is 
presented in Fig.\ref{fig4}c. The dependence on system parameters can be understood on the
basis of simple single-kick model. In this model the action of static imperfections in all
gates entering in one Grover iteration is replaced by a single kick unitary operator
$U_{eff}=\exp{(-i H_s n_g R)}$ acting after each iteration. Here $R$ is a dimensionless 
renormalization factor which takes into account that gates do not commute with $H_S$.
Figs.\ref{fig4}a,b show that this single kick approximation gives a good description of 
original averaged data with $R=0.56$. Thus, the renormalization effects play a significant role
and therefore this model does not describe the probability variation for a given
disorder realization. However, the averaged dependence is correctly reproduced.   

In the regime where the dynamics of Grover algorithm is dominated by four states subspace
(\ref{eq:4vect}) the single-kick model can be treated analytically. The matrix elements of the effective
Hamiltonian in this space are
\begin{equation}
H_{eff}=\left(
\begin{array}{cccc}
  A+a & 0 & -i\omega_G & 0 \\
  0 & A-a & 0 & -i\omega_G \\
  i\omega_G & 0 & B & b \\
  0 & i\omega_G & b &  B\\
\end{array}
\right),
\label{eq:Heff}
\end{equation}
where $A=-R n_g\sum_{i=1}^{n_q} a_i \langle\tau\vert\sigma_i^{(z)}\vert\tau\rangle$,
$B= R n_g\sum_{i<j}^{n_q} b_{i,j}-b$,  $a=-R n_g a_{n_q+1}$ and
$b=R n_g (b_{n_q+1,n_q+2-L_x}+b_{n_q+1,L_x}+b_{n_q,n_q+1}+b_{n_q+1-L_x,n_q+1})$
and qubits are arranged on $L_x\times L_y$ lattice, and numerated as $i=x+L_x (y-1)$, with
$x=1,\ldots,L_x$, $y=1,\ldots,L_y$. 
In the limit of large $n_q$ the terms $a,b$ are small compared to $A,B$ by a
factor $1/\sqrt{n_q}$ and $H_{eff}$ is reduced to $2\times 2$ matrix, which gives
$w_G=2\omega_G^2/[(A-B)^2+4\omega_G^2]$. For large $n_q$ the difference $A-B$ has a Gaussian
distribution with width $\sigma=R n_g \sqrt{n_q/3} \sqrt{\alpha^2+2\beta^2}=
\varepsilon R n_g \sqrt{n_q}$. The convolution
of $w_G$ with this distribution gives
\begin{equation}
\label{eq:2stMean}
\bar{w}_G=\sqrt{\pi/2}
(1-\operatorname{erf}(\sqrt{2}\omega_G/\sigma))\exp{(2\omega_G^2/\sigma^2)}\;\omega_G/\sigma
\end{equation}
This formula gives a good description of numerical data in Fig.\ref{fig4}c
that confirms the validity of single-kick model. 
For $\sigma \gg \omega_G$ and a typical disorder realization with $(A-B) \sim \sigma$ 
the actual frequency of Grover 
oscillations is strongly renormalized: $\omega \approx (A-B) \sim \sigma \gg \omega_G$, 
and in agreement with Fig.\ref{fig3} $\omega \sim \varepsilon/\varepsilon_c$.
In this typical case $w_G \sim \omega_G^2/\sigma^2 \ll 1$
(almost total probability is in the states $\vert \eta_0\rangle$,$\vert \eta_1\rangle$).
Hence, the total number
of quantum operations $N_{op}$, required for detection of searched state $\vert \tau\rangle$,
can be estimated as $ N_{op} \sim N_M/ \omega \sim \sigma/\omega_G^2 
\sim \varepsilon N/\varepsilon_c$, where
$N_M \sim 1/w_G \sim \sigma^2/\omega_G^2$ is a number of measurements required for detection
of searched state \cite{note}. Thus, in presence of strong static imperfections the
parametric efficiency gain of the Grover algorithm compared to classical one is of the
order $\varepsilon_c/\varepsilon$. For $\varepsilon \sim \omega_G$ the efficiency is comparable
with that of the ideal Grover algorithm while for $\varepsilon \sim \varepsilon_c$ there is no gain
compared to the classical case.

In summary, we have shown that the Grover algorithm remains robust against 
static imperfections inside a well defined domain
and determined the dependence of algorithm efficiency on the imperfection strength.

This work was supported in part by the EU IST-FET project EDIQIP and the NSA and ARDA under ARO
contract No. DAAD19-01-1-0553.


\begin{thebibliography}{99}
\bibitem{note} Here we consider only the subspace (\ref{eq:4vect}), 
a small probability leakage to all
other states is not crucial since it will be randomly distributed over $2N-4$ states.
\end{thebibliography}
\end{document}